\documentclass[prl,twocolumn,amssymb]{revtex4}
\usepackage{graphicx}
\usepackage{dcolumn}
\usepackage{bm}
\usepackage{wasysym}

\newcommand{\beq}{\begin{equation}}
\newcommand{\eeq}{\end{equation}}
\newcommand{\beqn}{\begin{eqnarray}}
\newcommand{\eeqn}{\end{eqnarray}}

\begin{document}
\title{Emergent Gravity at a Lifshitz Point from a Bose Liquid on the Lattice}
\author{Cenke Xu}
\affiliation{Department of Physics, Harvard University, Cambridge,
MA 02138, USA}
\author{Petr Ho\v{r}ava}
\affiliation{Berkeley Center for Theoretical Physics, University of
California, Berkeley, CA 94720, USA\\
Theoretical Physics Group, Lawrence Berkeley National
Laboratory, Berkeley, CA 94720, USA\\
Institute for the Physics and Mathematics of the Universe,
University of Tokyo, Kashiwa 277-8568, Japan}

\begin{abstract}

We propose a model with quantum bosons on the fcc
lattice, which has a stable algebraic Bose liquid phase at low energy.
We show that this phase is described by emergent quantum gravity
at the Gaussian $z = 3$ Lifshitz fixed point in $3+1$ dimensions.  The
stability of this algebraic Bose liquid phase is guaranteed by the gauge
symmetry of gravitons and self-duality of the low energy field theory.
By tuning one parameter in the lattice boson model we can drive a
phase transition between the $z=3$ Lifshitz gravity and another
algebraic Bose liquid phase, described by gravity at the $z=2$ Lifshitz
point.

\end{abstract}
\maketitle

The challenge of finding a satisfactory theory of quantum gravity has
stimulated theoretical physics for many decades.  We expect that such a
theory should make sense of the quantum fluctuations of the spacetime metric
at low-energy scales, while providing a quantum mechanical completion of the
system at short distances.  Recently, a new approach to this long-standing
puzzle has been proposed  \cite{horava2008,horava2009}.  In this approach,
gravity is treated using the traditional path integral methods of quantum
field theory, but without assuming Lorentz invariance as a fundamental
symmetry at short distances.  The gauge symmetries are those spacetime
diffeomorphisms that preserve a preferred foliation of spacetime by fixed
time slices, generated by
\beqn
\delta x^i=\xi^i(x^j,t),\quad\delta t=\zeta(t).\label{difff}
\eeqn
In terms of the spatial metric $g_{ij}(x^k,t)$, the shift vector $N_i(x^j,t)$
and the lapse function $N(t)$, the action is
\beqn S&=&\frac{1}{\kappa^2}\int
d^Dx\,dt\,N\sqrt{g}\,(L_K-L_V),\cr\cr L_K&=& \sum_{i,j}
K_{ij}K^{ij}-\lambda(\sum_i K_i^i)^2,\label{liflag} \eeqn where
\beqn K_{ij}=\frac{1}{2N}(\dot g_{ij}-D_iN_j-D_jN_i) \eeqn is the
extrinsic curvature of the constant time slices in the spacetime
foliation, $D_i$ is the covariant derivative defined by $g_{ij}$,
$\kappa$ and $\lambda$ are coupling constants, and the $L_V$ can
be an arbitrary local Lagrangian built from $g_{ij}$, its Riemann
tensor $R^i{}_{jk\ell}$, and the covariant derivative $D_i$,
without the use of time derivatives.

In this broader framework, new Gaussian fixed points of the renormalization
group (RG) are possible.  These novel fixed points are characterized by a
dynamical scaling exponent $z\neq 1$ (which depends on the choices made in
$L_V$), and they exhibit a Lifshitz-type scaling.
Given the possibility of such Lifshitz gravity fixed points, it is natural
to ask whether they can serve to provide an ultraviolet (UV) completion
of gravity, and whether the pattern of the RG flow can restore $z=1$
and Lorentz invariance in the infrared (IR) regime, where the theory can be
tested against general relativity.

In this paper, we show that this new class of Lifshitz gravity fixed points
\cite{horava2008,horava2009}
with $z=2$ and $z=3$ in $3+1$ dimensions can also emerge as {\it infrared\/}
fixed points, in the low-energy limit describing certain lattice systems.
The microscopic degrees of freedom of these systems are of the
conventional type familiar in condensed matter, but their collective behavior
leads at low energies to novel gapless phases, described by the Lifshitz
gravity fixed points.  The microscopic lattice degrees of freedom exhibit no
gauge symmetry:  The foliation-preserving gauge invariance at the Lifhitz
gravity fixed points is entirely emergent in the low-energy description of the
lattice system.  The more fundamental physical objects on short length scales
give rise to the Lifshitz gravitons as their low-energy and long-distance
collective excitations.

An analogous mechanism, with an emergent U(1) gauge symmetry, was realized
previously in the quantum dimer model on the cubic lattice.  In this model,
the short distance excitations are spin singlet valence bond fluctuations,
while the long length-scale excitations are photons
\cite{fradkin1990,sondhi2003}.  A similar photon phase can also be constructed
in spin models on the pyrochlore lattice \cite{hermele2004,wen2003}.  Given
these examples, it is natural to look for a lattice model whose infrared
behavior is controlled by the Lifshitz gravity fixed points of
\cite{horava2008,horava2009}.

It should be noted that there are two distinct ways in which one
can attempt to relate Lifshitz gravity to lattice models.  In this
paper, we work on a fixed rigid lattice, and find degrees of
freedom whose long distance dynamics is captured by the Gaussian
fixed points of Lifshitz gravity.  The lattice is nondynamical.
Alternatively, one can try to obtain Lifshitz gravity as a
continuum limit of a lattice model defined as a sum over a
suitable class of random triangulations of spacetime geometries.
Here it is the lattice itself that is dynamical, and no extra
degrees of freedom are invoked.  The most promising candidate for
such a random lattice model is offered by the causal dynamical
triangulations (CDT) approach to quantum gravity.  In the CDT
approach (see Ref.~\cite{ambjorn2009} for a review), a summation
over random lattices constrained to respect a preferred foliation
structure of spacetime serves as a nonperturbative definition of
quantum gravity, and yields a continuum limit with four
macroscopic spacetime dimensions at long distances.  It has been
suggested in \cite{horava2009a} (see also the recent paper
\cite{ambjorn2010}) that the CDT approach might be viewed as a
lattice regularization of Lifshitz gravity.  Further evidence for
this scenario comes from the qualitative behavior of the spectral
dimension of spacetime, which indicates that the model flows from
a $z=3$ UV fixed point to an $z=1$ fixed point at long distances
\cite{horava2009a}.

In condensed matter systems, gapless bosonic excitations are usually Goldstone
modes of certain spontaneously broken continuous symmetry.  For instance, the
phonon modes of solids are the Goldstone modes of translation symmetry, and
the magnons are the Goldstone modes of the broken spin rotation symmetry in
the magnetic ordered phase.  There has been a lot of interest recently in
finding bosonic phases with gapless excitations which do not originate from
breaking a symmetry, and which do not require fine tuning.  Such phases are
referred to as the {\it algebraic Bose liquid\/} (ABL) phases, because the
boson density correlation falls off algebraically in space and time.  An ABL
phase is defined and characterized by its low-energy field theory.  Searching
for stable ABL phases, or more generally Bose metal phases, is one of the
active fields in condensed matter theory
\cite{xu2006a,xu2006b,xufisher,motrunich2007,sheng2009,paramekanti2002}.
The photon phase of the quantum dimer model mentioned above is a
well-understood stable ABL phase which was first studied in high
energy community and then widely applied to condensed matter
systems, especially fractionalized states of strongly correlated
systems \cite{levin2009,sslee2005}.

Can we find other examples of ABL phases, especially in three spatial
dimensions?
In Ref.~\cite{xu2006a,xu2006b}, a novel ABL phase with $z=2$ dispersion was
proposed in a quantum boson model on the face centered cubic (fcc) lattice
with only local boson hopping and density repulsion. Although the microscopic
model only has the lattice point group symmetry and the U(1) global symmetry
corresponding to the conservation of the total boson number, a gauge
symmetry similar to linearized diffeomorphisms emerges at low energies.
We will review the low-energy effective field theory of this
model below, and show that it is given by the Lifshitz gravity of
Ref.~\cite{horava2008} at the $z=2$ Gaussian fixed point.   In the lattice
construction, the $z=2$ dispersion is protected by the gauge symmetry which
emerges at low energy and by the microscopic discrete symmetries of the
underlying degrees of freedom, and the stability of this ABL phase is
guaranteed by its self-duality.  In this work, we show that by
turning on one extra density repulsion term in the model proposed
in Ref.~\cite{xu2006a,xu2006b}, one can drive a phase transition
between the $z=2$ phase mentioned above and another stable $z=3$ phase.
We will show that this $z=3$ phase is also a stable ABL phase, with a
self-dual structure at low energy. The field theory of this ABL phase is
identical to the $z = 3$ Lifshitz gravity of Ref.~\cite{horava2009} at
the Gaussian fixed point.

\begin{figure}
\includegraphics[width=2.1in]{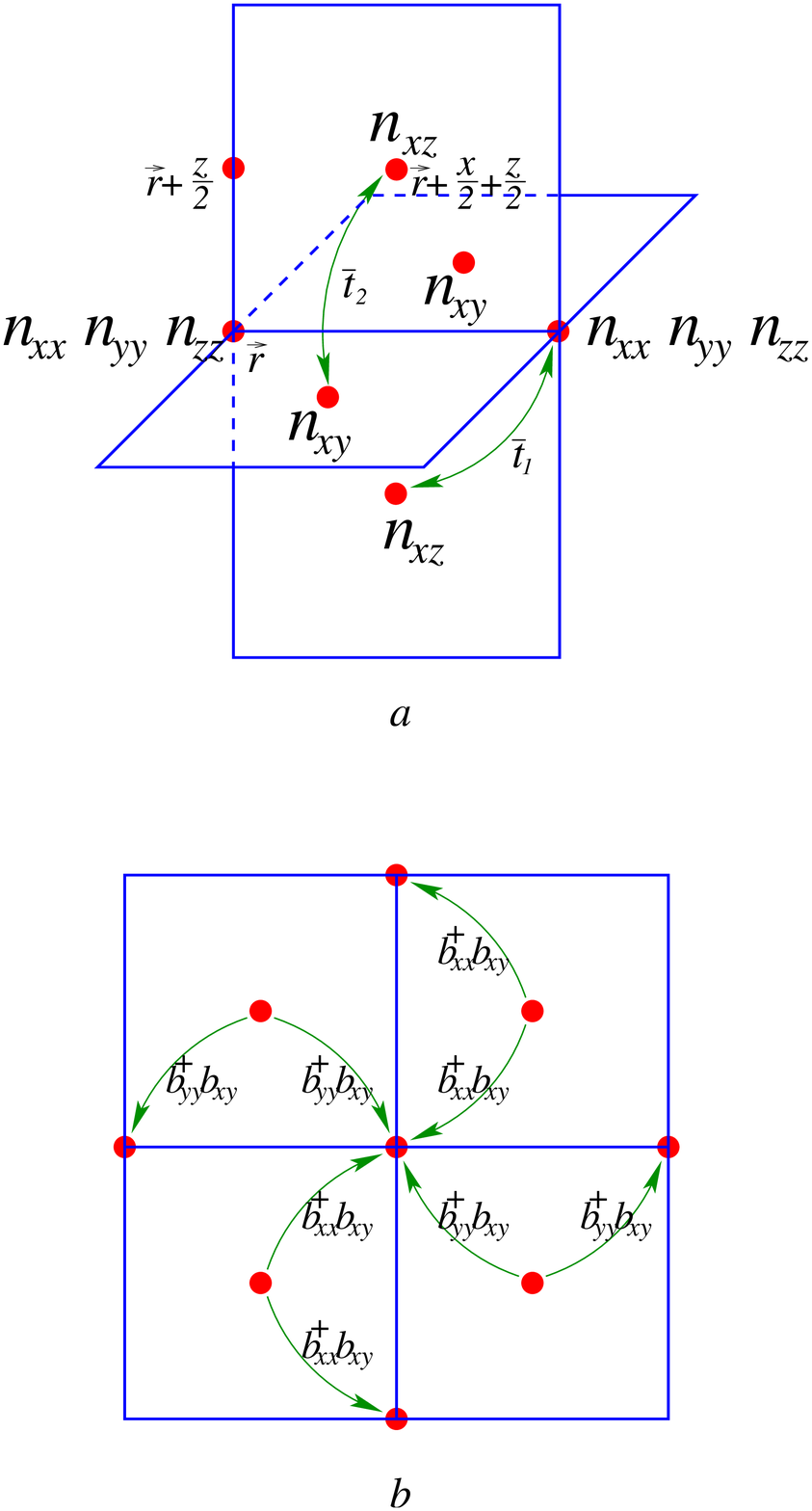}
\caption{$(a)$, The lattice structure of this model, with the
coordinates of site, link and plaquette. The term $H_{v, \
\hat{x}\ \mathrm{link}}$ in Eq.~(\ref{hamil}) involves all the
sites and plaquettes denoted with solid circles in this figure.
The $H_t$ term contains flavor dependent boson hopping between
sites and adjacent plaquettes (the $H_{sp}$ term in
Eq.~(\ref{hamil})), and also between nearest neighbor plaquettes
(the $H_{pp}$ term). $(b)$, the ring exchange term
$-t_1\cos(\mathcal{B}_{zz})$ in Eq.~(\ref{Hz2}), which includes
eight boson hoppings.} \label{z3gravity}
\end{figure}

We start with describing the full Hamiltonian of our lattice boson
model. This model is defined on the fcc lattice: The physical
quantities will be defined on both the sites and the unit square
faces of a cubic lattice. We denote each site of the cubic lattice
by $\vec{r} = (r_x, r_y, r_z)$, and each unit square in the
$(\hat{\imath},\hat{\jmath})$ plane by $\vec{r} \pm
\frac{\hat{\imath}}{2} \pm \frac{\hat{\jmath}}{2}$, with $i,j = x,
y, z$. As our dynamical variables, we assign three boson numbers
$(n_{xx}, n_{yy}, n_{zz})$ on each site of a cubic lattice, and
one boson number $n_{ij}$ to each face in
$\hat{\imath}\hat{\jmath}$ plane of the cubic lattice. The
corresponding creation and destruction operators will be denoted
by $b^\dagger_{ii},b_{ii},b^\dagger_{ij}$ and $b_{ij}$.  The
microscopic Hamiltonian contains the following terms, \beqn H &=&
H_t + H_v + H_u + H_{v^\prime}, \cr\cr H_t &=& - \bar{t}_1 H_{sp}
- \bar{t}_2 H_{pp}, \cr\cr H_{v, \ \hat{x}\ \mathrm{link}} &=&
V(2n_{xx, \vec{r}} + 2n_{xx,\vec{r} + \hat{x}} + n_{xy, \vec{r} +
\frac{\hat{x}}{2} + \frac{\hat{y}}{2}} + \cr\cr n_{xy, \vec{r} +
\frac{\hat{x}}{2} - \frac{\hat{y}}{2}} &+& n_{xz, \vec{r} +
\frac{\hat{x}}{2} + \frac{\hat{z}}{2}} + n_{xz, \vec{r} +
\frac{\hat{x}}{2} - \frac{\hat{z}}{2}} - 8\bar{n})^2, \cr\cr H_u =
\sum_{\vec{r}}\sum_{ii}&&\!\!\!\!\!\!\!\!\!\!\frac{u_1}{2}(n_{ii,
\vec{r}} - \bar{n})^2 + \sum_{i < j} \frac{u_2}{2}(n_{ij, \vec{r}
+ \frac{\hat{\imath}}{2} + \frac{\hat{\jmath}}{2}} - \bar{n})^2,
\cr\cr H_{v^\prime} &=& \sum_{\vec{r}} V^\prime(n_{xx,\vec{r}} +
n_{yy,\vec{r}} + n_{zz,\vec{r}} - 3\bar{n})^2. \label{hamil}\eeqn
$H_t$ include all the local boson hoppings. $H_{sp}$ is the
hopping between each site and its adjacent plaquettes,
$H_{sp} = \sum_{\vec{r}, i,j,k} b^\dagger_{kk, \vec{r}} b_{ij,
\vec{r} \pm \frac{\hat{\imath}}{2} \pm \frac{\hat{\jmath}}{2}} + h.c.$,
while $H_{pp}$ is the hopping between two nearest neighbor plaquettes
(Fig.~\ref{z3gravity}$a$). The exact values of the amplitudes
$\bar{t}_1, \bar{t}_2$ are unimportant, and we will tentatively
take both of them at the same order of magnitude $\bar{t}_1,
\bar{t}_2 \sim \bar{t}$. $H_v$ is a large density-density
interaction between bosons,  $H_{v, \ \hat{x}\ \mathrm{link}}$ in
Eq.~(\ref{hamil}) is the part of $H_v$ associated with one link
along the $\hat{x}$ direction; it is a sum of the boson numbers on
the two sites and the four faces sharing this link, and $\bar{n}$
is the average boson filling on each quantum state.  Contributions
to $H_v$ for links along $\hat{y}$ and $\hat{z}$ directions are
defined analogously.  $H_u$ is a small repulsive interaction
between each flavor of bosons on both sites and faces, with $u_1$
generally not equal to $u_2$. Finally, $H_{v^\prime}$ is another
on-site repulsive interaction on each site of the cubic lattice.
The lattice structure and the distribution of the boson number is
shown in Fig.~\ref{z3gravity}$a$.

If we first take $V^\prime = 0$, this model will reduce to the
model constructed in Ref.~\cite{xu2006a,xu2006b}. When $H_v$ is
dominant, it separates the Hilbert space into a high-energy
subspace and a low-energy subspace. The low-energy subspace is
subject to a local constraint: The sum in the bracket in $H_v$
vanishes for every link. If we define a symmetric tensor
$\mathcal{E}_{ij}$ as $\mathcal{E}_{ii, \vec{r}} =
-(-1)^{\vec{r}}2(n_{ii, \vec{r}} - \bar{n})$, $\mathcal{E}_{ij,
\vec{r} + \frac{\hat{\imath}}{2} + \frac{\hat{\jmath}}{2}} =
(-1)^{\vec{r}}(n_{ij, \vec{r} + \frac{\hat{\imath}}{2} +
\frac{\hat{\jmath}}{2}} - \bar{n})$ ($i \neq j$) this local
constraint can be compactly written as \beqn \sum_i
\nabla_i\mathcal{E}_{ij} = 0, \label{constraint}\eeqn Here and
throughout the paper, $\nabla_i$ denotes the lattice derivative:
$\nabla_i f(\vec{r}) = f(\vec{r} + \hat{\imath}) - f(\vec{r})$.
Eq.~(\ref{constraint}) simply states that $\mathcal{E}_{ij}$ is
covariantly conserved, or divergence-free.

The canonical conjugate variables of $n_{ij}$ on the lattice are
the phase angles $\theta_{ij}$ of the boson creation operators:
$b_{ij}\sim e^{- i\theta_{ij}}$.  They satisfy the commutation
relation $[n_{ab}, \theta_{cd}] = i\delta_{ac}\delta_{bd}$, $a
\leq b$, $c \leq d$. Using $\theta_{ij}$, we introduce a symmetric
tensor field $A_{ij}$ as $A_{ii,\vec{r}} =
-(-1)^{\vec{r}}\theta_{ii, \vec{r}}$ and $A_{ij, \vec{r} +
\frac{\hat{\imath}}{2} + \frac{\hat{\jmath}}{2}} = (-1)^{\vec{r}}\theta_{ij,
\vec{r} + \frac{\hat{\imath}}{2} + \frac{\hat{\jmath}}{2}}$ for $i
\neq j$. Under the discrete symmetries of time reversal
$\mathrm{T}$, lattice translations $\mathrm{T}_{\hat{k}}$ along
$\hat k$, and spatial reflection transformations
$\mathrm{P}_{\hat{k},s}$, the components of $A_{ij}$ transform as
\beqn \mathrm{T} &:& t \rightarrow -t, \ A_{ij} \rightarrow -
A_{ij}, \cr\cr \mathrm{T}_{\hat{k}} &:& \ \vec{r} \rightarrow
\vec{r} + \hat{k}, \ A_{ij} \rightarrow - A_{ij}, \cr\cr
\mathrm{P}_{\hat{k},s} &:& \ r_{\hat{k}} \rightarrow -
r_{\hat{k}}, \ A_{ik} \rightarrow - A_{ik}, \ i \neq k, \cr\cr &&
A_{ij} \rightarrow A_{ij}, \ i \neq k, j\neq k, \cr\cr && A_{ii}
\rightarrow A_{ii}. \label{dissy}\eeqn Notice that we define
$\mathrm{P}_{\hat{k},s}$ to be the site-centered reflection of the
lattice, where the origin is located at one of the sites of the
cubic lattice. The transformation of $\mathcal{E}_{ij}$ is almost
the same as $A_{ij}$, except that $\mathcal{E}_{ij}$ is even under
time reversal.

The local constraint on $\mathcal{E}_{ij}$ Eq.~(\ref{constraint})
can be interpreted as a Gauss constraint, associated with a partially
fixed gauge symmetry.  It will generate the following gauge symmetries on
$A_{ij}$: \beqn A_{ij} \rightarrow A_{ij} + \nabla_i f_j + \nabla_j f_i.
\label{gaugez2}\eeqn Of course, this gauge symmetry was absent in
the microscopic Hamiltonian Eq.~(\ref{hamil}):  It only emerges in
the low-energy Hilbert subspace with constraints imposed by $H_v$.
This mechanism is analogous to the way in which the emergent U(1) gauge
symmetry appears in the photon phase of the 3d quantum dimer models
\cite{fradkin1990,sondhi2003} and similar spin models
\cite{hermele2004,wen2003}. If we interpret $A_{ij}$
as small fluctuations of the metric tensor $g_{ij}$ around the flat
background,
\beqn
g_{ij}\approx\delta_{ij}+A_{ij},\label{linmet}
\eeqn
the emergent gauge symmetry in Eq.~(\ref{gaugez2}) corresponds to the
linearized form of the foliation-preserving diffeomorphisms Eq.~(\ref{difff})
of Lifshitz gravity, with $\xi^i(x,t)\approx f_i$.

The low-energy dynamics of this system has to be invariant under
the gauge transformation Eq.~(\ref{gaugez2}). The lowest order
gauge invariant dynamics is generated at the eighth order
perburbation of $\bar{t}/V$, which can be written as \beqn
H_{\mathrm{eff}} &=& \sum_{\vec{r}} - \sum_{i \neq j}
t_1\cos(R_{ijij})_{\vec{r}} -\!\!\!\!\!\!\! \sum_{i\neq j, j\neq
k, k\neq i}\!\!\!\! t_2\cos(R_{ijik})_{\vec{r}}, \cr\cr R_{ijij}
&=& \sum_{a,b,c,d,k} -
\epsilon_{ijk}\epsilon_{kab}\epsilon_{kcd}\nabla_a\nabla_cA_{bd},
\ i \neq j, \cr\cr R_{ijik} &=& \sum_{a,b,c,d}
\epsilon_{jab}\epsilon_{kcd}\nabla_a\nabla_cA_{bd}, \ i\neq j,
j\neq k, k\neq i. \eeqn With our identification of $A_{ij}$ as the
metric fluctuations around the flat metric, $R_{ijij}$ and
$R_{ijik}$ represent the six independent components of the
linearized curvature tensor of $g_{ij}$.  It is convenient to
introduce another rank-two symmetric tensor $\mathcal{B}_{ij}$ as
$\mathcal{B}_{ii} = \frac{1}{2}\epsilon_{ijk}R_{jkjk}$ and
$\mathcal{B}_{ij} = - R_{ikjk}$. $\mathcal{B}_{ij}$ defined this
way is also covariantly constant : $\nabla_i\mathcal{B}_{ij} = 0$.
The transformation of $\mathcal{B}_{ij}$ under the discrete
lattice symmetries of Eq.~(\ref{dissy}) is the same as for
$A_{ij}$. The definition of $\mathcal{B}_{ij}$ on the lattice is
given in the appendix. In geometric terms, $\mathcal{B}_{ij}$ are
simply the components of the linearized Einstein tensor
$R_{ij}-\frac{1}{2}Rg_{ij}$ of the metric (\ref{linmet}).

Now the full low-energy Hamiltonian reads \beqn H = H_u +
H_{\mathrm{eff}} = \sum_{\vec{r}}\sum_{ii} \frac{u_1}{8}
\mathcal{E}_{ii,\vec{r}}^2 + \sum_{i < j} \frac{u_2}{2}
\mathcal{E}_{ij, \vec{r}}^2 \cr\cr - \sum_{i}
t_1\cos(\mathcal{B}_{ii})_{\vec{r}} - \sum_{i\neq j}
t_2\cos(\mathcal{B}_{ij})_{\vec{r}}. \label{Hz2}\eeqn $t_1$ and
$t_2$ are both at the order of $\sim \bar{t}^8/V^7$. Physically
$t_1$ and $t_2$ terms correspond to high order hoppings of bosons.
For instance $t_1$ term stands for the eighth order hopping
process depicted in Ref.~\cite{xu2006a,xu2006b} and and
Fig.~\ref{z3gravity}$b$. We want to emphasize that the Hamiltonian
Eq.~(\ref{hamil}) is not fine-tuned, in the sense that one is
allowed to turn on small local perturbations of any kind that are
compatible with the global symmetry of Eq.~(\ref{hamil}), and the
low-energy Hamiltonian will always take the same form as
Eq.~(\ref{Hz2}).

The Hamiltonian in Eq.~(\ref{Hz2}) has a continuum limit Gaussian field
theory, which characterizes an ABL phase. In the field theory we
replace the lattice derivative $\nabla_i$ by the continuum limit
derivative $\partial_i$, and expand the cosines in
$H_{\mathrm{eff}}$ at its minima $- \cos(\mathcal{B}_{ij}) \sim
\mathcal{B}_{ij}^2/2$ to the leading nontrivial order (a spin-wave
expansion).  After we replace $\mathcal{E}_{ij}$ by
$\mathcal{E}_{ij}\sim\dot{A}_{ij}$, the resulting Gaussian field
theory is described by the following Lagrangian,
 \beqn \mathcal{L} = \sum_i
\frac{1}{2u_1}(\dot{A}_{ii})^2 +
\frac{t_1}{2}\mathcal{B}^2_{ii} + \sum_{i < j}
\frac{1}{2u_2}(\dot{A}_{ij})^2 +
\frac{t_2}{2}\mathcal{B}^2_{ij}.~\label{lagz2} \eeqn
The dispersion of the collective modes can be solved
straightforwardly, and the result is quadratic, implying $z=2$.
Importantly, the linear dispersion is ruled out by the gauge symmetry and
the lattice symmetries; no other terms more relevant than those in
Eq.~(\ref{lagz2}) are allowed. The terms $\sum_{i\neq j}\mathcal{E}_{ii}
\mathcal{E}_{jj}$ and $\sum_{i\neq j}\mathcal{B}_{ii}\mathcal{B}_{jj}$
are in principle allowed by the symmetry of the system, but it will
not change the $z=2$ dispersion. For instance, $\sum_{i\neq j}\mathcal{E}_{ii}
\mathcal{E}_{jj}$ term can be induced with the $V^\prime$ term in
Eq.~(\ref{hamil}), while $\sum_{i\neq j}\mathcal{B}_{ii}\mathcal{B}_{jj}$
can be induced with even higher order boson ring exchange.

It is now easy to see that the theory in Eq.~(\ref{lagz2}),
together with the constraint of Eq.~(\ref{constraint}), is
equivalent to the Gaussian limit of the $z=2$ Lifshitz quantum
gravity proposed in \cite{horava2008,horava2009}. Indeed, setting
$N_i=0$ as our gauge choice in Lifshitz gravity yields
Eq.~(\ref{constraint}) as the equation of motion, from varying
$N_i$ in the action given in Eq.~(\ref{liflag}).  Furthermore, the
most general potential term $S_V$ in the $z=2$ Lifshitz gravity
Lagrangian in $3+1$ dimensions is a sum of two terms, \beqn S_V =
\sum_{i,j} \kappa_1R_{ij}R^{ij}+\kappa_2 R^2,\label{z2sv}\eeqn
while the kinetic term $S_K$ takes the form in Eq.~(\ref{liflag}).
One difference between the lattice system and the field theory
Eq.~(\ref{z2sv}) is that, the lattice system does not have the
O(3) spatial rotation symmetry, therefore the low-energy field
theory deduced from our lattice model does not automatically
acquire this O(3) symmetry.  However, one can tune the parameters in
Eq.~(\ref{hamil}) to restore the O(3) symmetry in the low-energy
limit, and then the field theory will be identical to
Eq.~(\ref{liflag}) plus $S_V$.  Interestingly, if we tune the microscopic
parameters to achieve the O(3) symmetry, the ABL phase picks out the
special case of the Lifshitz gravity models, satisfying the additional
property of detailed balance in the sense of \cite{horava2008,horava2009},
with a fixed value of the coupling $\lambda$ in Eq.~(\ref{liflag}).
Extending the relationship beyond this simplest case requires that
the additional terms $\sum_{i\neq j}\mathcal{B}_{ii}\mathcal{B}_{jj}$ and
$\sum_{i\neq j}\mathcal{E}_{ii} \mathcal{E}_{jj}$ mentioned above are also
generated.

Having established the map bewteen the low-energy continuum Hamiltonian of
the lattice boson model and the Hamiltonian of $z=2$ Lifshitz gravity, a few
comments are in order:

(i) In Lifshitz gravity with the full foliation-preserving
diffeomorphism invariance of Eq.~(\ref{difff}), one additional
global constraint follows from the variation of $N(t)$ in the
action Eq.~(\ref{liflag}).  Since $N(t)$ is only a function of
time, this yields an {\it integral\/} constraint, equivalent to
the vanishing of the total Hamiltonian on physical states. Not
imposing the $\delta t=\zeta(t)$ invariance as a gauge symmetry
would eliminate this Hamiltonian constraint.  In the lattice
approach, the integral Hamiltonian constraint does not seem to be
necessary for self-consistency; however, we could consider
imposing it in addition to the local constraint given in
Eq.~(\ref{constraint}), thus reproducing the full set of
symmetries of the minimal version of Lifshitz gravity.

(ii) In \cite{horava2008,horava2009}, a fully interacting
nonlinear version of Lifshitz gravity has been proposed, and it
was argued that models with $z>1$ naturally flow in the IR to
$z=1$, under the influence of relevant terms in $S_V$.  The most
relevant terms are $R$ -- the Einstein-Hilbert term responsible
for $z=1$, and $\Lambda$ -- the cosmological constant term. In the
self-interacting Lifshitz gravity theory, such terms are always
expected to be generated under RG\@, and one might wonder why they
do not automatically arise in the lattice model. The resolution of
this puzzle is simple:  The microscopic structure of our lattice
model implies that at the $z=2$ Gaussian fixed point, the discrete
symmetries of Eq.~(\ref{dissy}) will hold.  It turns out that
these symmetries not only prevent the relevant terms $R$ and
$\Lambda$ from being generated, they are also incompatible with
turning on the self-interaction coupling of the full nonlinear
Lifshitz gravity.

(iii) One should be more careful with the naive expansion of the
cosine functions appearing in Eq.~(\ref{Hz2}).  The way we
constructed the low-energy Hamiltonian from the microscopic
degrees of freedom, the $A_{ij}$ variables are compact, with
radius $2\pi$.  This compactification does not have a direct
analogy in Lifshitz gravity.  Moreover, it can lead to topological
excitations, which in turn cause tunnelling between different
minima of the cosine functions in Eq.~(\ref{hamil}). Just like in
the case of compact QED in $2+1$ dimensions, the compactification
has the potential to destroy the gapless excitations when it is
relevant.

The relevance or irrelevance of the compactification and
topological excitations can be most naturally studied in the dual
picture. The low-energy Hamiltonian of this ABL phase can be
schematically written as $H = \mathcal{E}^2 + \mathcal{B}^2$,
$\mathcal{E}$ and $\mathcal{B}$ are both symmetric,  covariantly
constant tensors with six independent components, which suggests a
self-dual structure exchanging $\mathcal{E}$ and $\mathcal{B}$.
The dual field $h_{ij}$ and the dual momentum $\pi_{ij}$ are
defined as \beqn \mathcal{E}_{ij} =
\epsilon_{iab}\epsilon_{jcd}\nabla_a\nabla_ch_{bd}, \
\mathcal{B}_{ii} = 2 \pi_{ii}, \ \mathcal{B}_{ij} = \pi_{ij} (i
\neq j).\eeqn Notice that all the derivatives are lattice
derivatives, and $h_{ij}$ and $\pi_{ij}$ are defined on the sites
and plaquettes of the cubic lattice (for details, see
the appendix). One can check the commutator and verify that
$h_{ij}$ and $\pi_{ij}$ are canonical conjugate variables,
$[\pi_{ij}, h_{kl}] = i\delta_{ik}\delta_{jl}$ $i \leq j$, $k \leq
l$, and we can replace $\pi_{ij}$ by $\dot{h}_{ij}$. Under this
duality transformation, $\mathcal{E}_{ij}$ and $\mathcal{B}_{ij}$
are exchanged, and the dual low-energy continuum limit field
theory is precisely the same as the original model.  Therefore,
this ABL phase is self-dual with the dual gauge symmetry $h_{ij}
\rightarrow h_{ij} +
\partial_i\tilde f_j + \partial_j\tilde f_i$ in the continuum
limit. Due to the compactness of $A_{ij}$, $\mathcal{E}_{ij}$ and
$h_{ij}$ both take discrete eigenvalues, therefore at the
microscopic level the dual Lagrangian allows for the periodic
potential $\hat{O}_v \sim \cos(2\pi h_{ij})$ which we refer to as
the vertex operator. However, this vertex operator violates the
dual continuum limit gauge symmetry of the ABL phase, and hence
the correlation function between $\hat{O}_v$ vanishes at distance
larger than $V/\bar{t}$. Thus, $\hat{O}_v$ is irrelevant at the
ABL gaussian fixed point.

The degrees of freedom in the $z=2$ ABL phase indeed match the
count of degrees of freedom in $z=2$ Lifshitz gravity \cite{horava2008}:
In addition to the two transverse-traceless polarizations of the
graviton, there is an additional ``scalar graviton'' mode.  In the
lattice model, these three polarizations appear as independent
collective modes of the Hamiltonian in Eq.~(\ref{Hz2}). The scalar
mode corresponds to the scale factor of the spatial metric. It was
shown in \cite{horava2009} that in classical Lifshitz gravity, the
scalar graviton can be eliminated by extending the gauge
invariance to include an anisotropic Weyl symmetry, introduced
first in \cite{horava2008} (and further studied in
\cite{hmt2009}). In $3+1$ dimensions, this requires the dynamical
exponent to be $ z = 3 $, freezes the coupling constant $\lambda$
in the action (\ref{liflag}) to be $\lambda=1/3$, and requires
that the spatial part $L_V$ of the action be conformally
invariant.

At the microscopic level of our lattice model, the elimination of
the scalar graviton can be arranged simply by turning on
$H_{v^\prime}$ in Eq.~(\ref{hamil}),  When both $H_v$ and
$H_{v^\prime}$ are dominant, the constraint on $\mathcal{E}_{ij}$
in the low-energy subspace becomes \beqn \sum_i
\nabla_i\mathcal{E}_{ij} = 0, \quad \sum_i \mathcal{E}_{ii} = 0.
\eeqn Thus, $\mathcal{E}_{ij}$ is now not only symmetric and
covariantly constant, but also traceless. This tracelessness
constraint can be interpreted as the Gauss constraint associated
with gauge fixing of another gauge symmetry, which acts on
$A_{ij}$ via
 \beqn A_{ij} \rightarrow A_{ij} +
\delta_{ij}\varphi, \label{gaugez3}\eeqn where $\varphi$ is an arbitrary
scalar field. Now the trace of $A_{ij}$ becomes an
unphysical gauge degree of freedom, so there are only two gapless
modes at low energy.

The new gauge symmetry in Eq.~(\ref{gaugez3}) will turn out to be
the anisotropic Weyl invariance of Lifshitz gravity mentioned
above, here in the linearized approximation around the Gaussian
$z=3$ fixed point.  In order to see that the new constraint is
forcing the model to $z=3$ at low energies, note that the low
energy Hamiltonian is also modified by turning on $V'$. Let us
assume the dynamical term can be written as \beqn
H_{\mathrm{eff}^\prime} = \sum_{\vec{r}} \sum_{i} -
t_3\cos(a\mathcal{C}_{ii})_{\vec{r}} - \sum_{i\neq
j}t_4\cos(b\mathcal{C}_{ij})_{\vec{r}}, \eeqn where
$\mathcal{C}_{ij}(A_{kl})$ is a linear functional of $A_{kl}$.
This new tensor $\mathcal{C}_{ij}$ must be invariant under
the gauge transformations of Eq.~(\ref{gaugez2}) as well as
Eq.~(\ref{gaugez3}). The tensor of the lowest order in derivatives
satisfying this requirement is the (linearized form of the) Cotton
tensor \cite{horava2008,horava2009}: \beqn \mathcal{C}_{ij} =
\epsilon_{ikl}\nabla_k(R_{jl} - \frac{1}{4}R\delta_{jl}).
\label{cotton}\eeqn Here $R_{jl} = \sum_k R_{jklk}$, $R = \sum_j
R_{jj}$ are the linearized Ricci curvature and scalar curvature
respectively. Under the lattice symmetries of Eq.~(\ref{dissy}),
the Cotton tensor transforms as \beqn \mathrm{T} &:&
\mathcal{C}_{ij} \rightarrow -\mathcal{C}_{ij}, \cr\cr
\mathrm{T}_{\hat{k}} &:& \mathcal{C}_{ij} \rightarrow
-\mathcal{C}_{ij}, \cr\cr \mathrm{P}_{\hat{k},s} &:&
\mathcal{C}_{ik} \rightarrow \mathcal{C}_{ik}, \ i \neq k, \cr\cr
&& \mathcal{C}_{ij} \rightarrow - \mathcal{C}_{ij}, \ i \neq k,
j\neq k, \cr\cr && \mathcal{C}_{ii} \rightarrow -
\mathcal{C}_{ii}. \eeqn One can straightforwardly verify that
$\mathcal{C}_{ij}$ is gauge invariant, symmetric, traceless and
covariantly constant: \beqn \mathcal{C}_{ij} = \mathcal{C}_{ji}, \
\sum_i \mathcal{C}_{ii} = 0, \ \sum_i \nabla_i\mathcal{C}_{ij} =
0. \eeqn The definition of $\mathcal{C}_{ij}$ on the lattice is
given in the appendix. From our microscopic Hamiltonian, the
effective low-energy term $H_{\mathrm{eff}^\prime}$ can only
emerge at fairly high order of perturbation. For instance, the
$t_3$ term can be generated at order 32 in $\bar{t}/V$. Now the
full low-energy Hamiltonian reads \beqn H = H_u + H_{\mathrm{eff}}
= \sum_{\vec{r}}\sum_{ii} \frac{u_1}{2} \mathcal{E}_{ii,\vec{r}}^2
+ \sum_{i < j} \frac{u_2}{2} \mathcal{E}_{ij, \vec{r}}^2 \cr\cr -
\sum_{i} t_3\cos(\mathcal{C}_{ii})_{\vec{r}} - \sum_{i\neq j}
t_4\cos(2\mathcal{C}_{ij})_{\vec{r}}. \label{Hz3}\eeqn Since
$\mathcal{C}_{ij}$ involves three spatial derivatives, after the
spin-wave expansion $- \cos(b\mathcal{C}_{ij}) \sim b^2
\mathcal{C}_{ij}^2/2$, the ABL phase which is described by the
continuum limit Gaussian field theory of Hamiltonian in
Eq.~(\ref{Hz3}) has collective excitations with the $z=3$
dispersion.
Notice that the lattice symmetry does not require $t_3 = t_4$, but
the ratio $t_3/t_4$ can be tuned by the ratio
$\bar{t}_1/\bar{t}_2$ on the lattice.

In \cite{horava2009}, a
nonlinear self-interacing $z=3$ Lifshitz gravity has been
constructed.  In this construction, the nonlinear Cotton tensor
plays a central role, with
$S_V\propto\mathcal{C}_{ij}\mathcal{C}^{ij}$.  When $t_3 = t_4$,
the low-energy Hamiltonian of the ABL phase is identical to the
Gaussian point of the $z=3$ Lifshitz gravity of \cite{horava2009}.
This emergent theory is gauge invariant not only under linearized
foliation-preserving diffeomorphisms, but also under the $z=3$
version of the anisotropic Weyl transformation
\cite{horava2008,horava2009,hmt2009}.

Are the gapless $z=3$ modes of this model stable when $A_{ij}$ are
treated as compact variables? To address this question, it is
again convenient to go to the dual formalism. The continuum limit
Hamiltonian takes the schematic form $H = \mathcal{E}^2 +
\mathcal{C}^2$, and the fact that $\mathcal{E}_{ij}$ and
$\mathcal{C}_{ij}$ are both symmetric, traceless and covariant
tensors strongly suggests this theory also has a self-dual
structure. To prove the self-duality, we define dual variables
$\tilde{h}_{ij}$ and $\tilde{\pi}_{ij}$ as \beqn \mathcal{E}_{ij}
= \mathcal{C}_{ij}(\tilde{h}_{kl}), \ \tilde{\pi}_{ii} =
\frac{1}{2}\mathcal{C}_{ii}(A_{kl}), \ \tilde{\pi}_{ij} =
\mathcal{C}_{ij}(A_{kl}),\ i\neq j. \eeqn Here we have treated
$\mathcal{C}_{ij}$ as a linear functional of the tensor field $A_{ij}$
or $\tilde{h}_{ij}$. Unlike $h_{ij}$ defined previously,
$\tilde{h}_{ii}$ are defined on the dual lattice sites, which are
the centers of the unit cubes, and $\tilde{h}_{ij}$ are defined on
the dual plaquettes, which are the links of the original lattice
(for details, see the appendix). For instance, a link
along the $\hat{z}$ direction is a dual plaquette in the $\hat{x}\hat{y}$
plane. Again, one can straightforwardly verify that
$\tilde{h}_{ij}$ and $\tilde{\pi}_{ij}$ are a pair of conjugate
variables, and $\mathcal{E}_{ij}$ and $\mathcal{C}_{ij}$ are
exchanged under duality. The dual graviton $\tilde{h}_{ij}$ of
this ABL phase enjoys the same gauge symmetry as $A_{ij}$, hence
the vertex operator $\hat{O}_v \sim \cos(2\pi\tilde{h}_{ij})$ is
irrelevant. Therefore the $z = 3$ ABL phase is also a stable
gapless phase.

This self-duality structure completes the argument of the
stability of both the $z = 2$ and $z = 3$ ABL phases in this
model. The self-duality can also be proved in the Euclidean
space-time, in the same way as the duality of ordinary classical
statistical mechanics models \cite{kogut}. If the dual theory of a
lattice gauge model did not have a large enough gauge symmetry,
one would have to fine-tune the system to get a gapless ABL state.
The most well-understood example is the compact QED in $2+1$
dimensions, where the dual theory is a U(1) rotor model without
gauge symmetry.  In that case, the vertex operator which
corresponds to the monopoles in spacetime will destroy the gapless
photon phase. Another example studied recently is the quantum
plaquette model with gauge symmetry $A_{ij} \rightarrow
\nabla_i\nabla_j \varphi$. The dual theory of the plaquette model
does have a gauge symmetry, but this symmetry is not strong enough
to protect the gapless ABL phase \cite{xuwu2008}.

In Ref.~\cite{guwen2009}, graviton-like collective modes were also
obtained through a quantum boson model on the lattice, but the
relation of such lattice models to Lifshitz gravity was not
noticed.  It is worth emphasizing that in our context, the
Lifshitz-type graviton Hamiltonian in Eq.~(\ref{Hz3}) can be
derived {\it without any fine tuning\/} from a simple boson model
of Eq.~(\ref{hamil}) through perturbation theory, and both the
$z=2$ and $z=3$ ABL phase are stable due to their self-dual
nature.

Besides the difference in the dispersion, the $z = 2$ and $z=3$
phases also have different algebraic correlations. For instance,
in the $z = 2$ phase the equal-time density fluctuation
correlation falls off as \beqn \langle \delta n(0)\delta
n(\vec{r})\rangle \sim \langle \mathcal{B}(\tilde{h})_0
\mathcal{B}(\tilde{h})_{\vec{r}} \rangle \sim \frac{1}{r^5}, \eeqn
while for the $z = 3$ phase this correlation falls off as \beqn
\langle \delta n(0)\delta n(\vec{r})\rangle \sim \langle
\mathcal{C}(\tilde{h})_0 \mathcal{C}(\tilde{h})_{\vec{r}} \rangle
\sim \frac{1}{r^6}. \eeqn (In these two equations, the flavor
dependence of the correlation has been ignored for simplicity.)

In the original Hamiltonian, the $z=2$ and $z=3$ phases were
obtained by dialing a small versus large value of the coupling
$V'$ in $H_{v^\prime}$. As a result, by increasing $V^\prime$ one
can drive a phase transition between these two ABL phases with
different dynamical scalings. Compared to the $z=2$ phase, the
$z=3$ phase has one extra gauge symmetry, of anisotropic Weyl
transformations (\ref{gaugez3}).  Therefore, if we start with the
$z = 3$ phase, this phase transition will be reminiscent of the
Higgs transition.  In the quantum dimer model, the U(1) gauge
symmetry is Higgsed by condensing the dimer vacancies that carry
U(1) gauge charge.  The phase transition can be described by the
rotor Hamiltonian $H_r = - t \cos(\vec{\nabla}\theta - \vec{A})$
\cite{balentsdimer}, where $\theta$ is the phase angle of the
dimer vacancy creation operator $\psi \sim e^{i\theta}$. After the
condensation of $\psi$, the gauge field $A_\mu$ acquires a
longitudinal mode by absorbing the Goldstone mode of $\psi$, which
makes the gauge field an ordinary gapped vector field.

In our case, the transition between the two ABL phases can also be
intuitively described as closing the gap of the trace mode of
$\mathcal{E}_{ij}$, which can be described by condensing $\varphi$
in Eq.~(\ref{gaugez3}).  Since all the gauge symmetries in
Eq.~(\ref{gaugez2}) are still preserved after the transition, the
condensate of $\varphi$ should not violate Eq.~(\ref{gaugez2}).
With these observations, this transition can be described by the
following Lagrangian, \beqn L = \frac{1}{\gamma}\dot\varphi^2 -
\sum_{i\neq j} t_1\cos(\nabla_i^2 \varphi + \nabla_j^2 \varphi -
R_{ijij}) \cr\cr - \sum_{i\neq j, j\neq k, k\neq i} t_2
\cos(\nabla_j\nabla_k\varphi - R_{ijik}) + \cdots
\label{transition}\eeqn The condensation of $\varphi$ changes the
spectrum from two $z=3$ modes of $A_{ij}$ and one gapped mode of
$\varphi$ to three $z=2$ modes of $A_{ij}$, in a generalization of
the ordinary Higgs transition to the scalar mode of Lifshitz
gravity. When $\varphi$ is ordered, this Lagrangian restores the
ring exchange terms in Eq.~(\ref{Hz2}) of the $z = 2$ phase; when
$\varphi$ is disordered, after integrating out $\varphi$ one
should recover the phase which is invariant under the
transformations in Eq.~(\ref{gaugez2}) and Eq.~(\ref{gaugez3}).
The $z = 3$ ABL phase is then the only candidate. From the
condensed matter perspective, the phase transition in
Eq.~(\ref{transition}) is beyond the ordinary Ginzburg-Landau
paradigm, because neither one of the two phases can be
characterized by a local order parameter. More thorough RG studies
for Eq.~(\ref{transition}) are required to determine the nature of
this transition.

One interesting question to ask is whether we can obtain
relativistic gravitons, with a linear dispersion, from the
lattice. In Ref.~\cite{xu2006b}, it was proposed that a long-range
interaction can change the dispersion to $z=1$, but a local theory
leading to $z=1$ gravitons at long distances is still unavailable.
As was discussed in Ref.~\cite{xu2006b} and Ref.~\cite{guwen2006},
a Chern-Simons like term $A_{ij}\mathcal{B}_{ij}$ can lead to a
linear dispersion, but this term is only gauge invariant up to a
boundary term; therefore, it cannot be generated in the same way
as the $t_1$ and $t_2$ terms in Eq.~(\ref{Hz2}) through
perturbation theory. One possibility is to generate this term by
coupling the graviton field $A_{ij}$ to a matter field with a
gapless boundary state, just like the CS term for a U(1) gauge
field can be generated by coupling the gauge field to a massive
Dirac fermion with edge states.

In addition to the linear dispersion, another meaningful goal is
to obtain the full nonlinear Lifshitz gravity of self-interacting
gravitons, instead of the linearized theory at the Gaussian fixed
point. As we pointed out in our comment (ii) below
Eq.~(\ref{z2sv}), this difficulty is intimately related to the
existence of the discrete symmetries in Eq.~(\ref{dissy}), implied by the
microscopic dynamics of the lattice model.
These symmetries do not allow the natural self-interaction coupling of
Lifshitz gravity \cite{horava2008,horava2009} to be turned on.
It is an interesting challenge to see if our framework
can be extended so that its discrete symmetries no longer prevent the
self-interaction of gravitons.  Note that the discrete
symmetries of Eq.~(\ref{dissy}) act naturally on the $A_{ij}$ variables
representing the fluctuations around the fixed flat
background, but they do not appear to have a geometrically natural extension
to the full metric $g_{ij}$.  Thus, they are associated with the fixed flat
spatial geometry, here represented by the fixed flat fcc lattice.
These background-dependent discrete symmetries indeed played an important
role in our construction of Lifshitz gravity from a lattice system: They
prevented the Einstein-Hilbert term and the cosmological constant term from
being generated, allowing the $z=2$ and $z=3$ ABL phases to be stable at low
energies.  Since the full nonlinear Lifshitz gravity of
\cite{horava2008,horava2009} does not require a choice of a preferred flat
background, it is natural to speculate that attempts to turn on the
self-interaction of gravitons in our lattice framework may ask for the
underlying lattice itself to become dynamical.
We will leave these topics to future studies.

We wish to thank the organizers of the KITP Miniprogram on
{\it Quantum Criticality and AdS/CFT Correspondence}: Sean Hartnoll,
Joe Polchinski and Subir Sachdev, for their hospitality in Santa Barbara in
during an important stage of this work in July 2009\@.
P.H. has been supported
by NSF Grants  PHY-0555662 and PHY-0855653, DOE Grant DE-AC02-05CH11231,
and by the Berkeley Center for Theoretical Physics.

\section{Appendix: tensor fields on the lattice}

On the lattice, $\pi_{zz}$ on site $\vec{r}$ is defined as \beqn
2\pi_{zz, \vec{r}} &=& (-1)^{\vec{r}}(\theta_{yy, \vec{r} +
\hat{x}} + 2 \theta_{yy, \vec{r}} + \theta_{yy, \vec{r} - \hat{x}}
\cr\cr && + \theta_{xx, \vec{r} + \hat{y}} + 2 \theta_{xx,
\vec{r}} + \theta_{xx, \vec{r} - \hat{y}} \cr\cr && - 2\theta_{xy,
\vec{r} + \frac{\hat{x}}{2} + \frac{\hat{y}}{2}} - 2\theta_{xy,
\vec{r} - \frac{\hat{x}}{2} + \frac{\hat{y}}{2}} \cr\cr && -
2\theta_{xy, \vec{r} + \frac{\hat{x}}{2} - \frac{\hat{y}}{2}} -
2\theta_{xy, \vec{r} - \frac{\hat{x}}{2} - \frac{\hat{y}}{2}}) =
\mathcal{B}_{zz}. \eeqn The ring exchange $\cos(\mathcal{B}_{zz})$
corresponds to the high order of boson hopping depicted in
Fig.~\ref{z3gravity}$b$. $\pi_{yz}$ on plaquette $\vec{r} +
\frac{\hat{y}}{2} + \frac{\hat{z}}{2}$ reads \beqn \pi_{yz,
\vec{r} + \frac{\hat{y}}{2} + \frac{\hat{z}}{2}} &=&
(-1)^{\vec{r}} \times \cr\cr &(& \theta_{yz, \vec{r} + \hat{x} +
\frac{\hat{y}}{2} + \frac{\hat{z}}{2} } + 2 \theta_{yz, \vec{r} +
\frac{\hat{y}}{2} + \frac{\hat{z}}{2} } + \theta_{yz, \vec{r} -
\hat{x} + \frac{\hat{y}}{2} + \frac{\hat{z}}{2}} \cr\cr & + &
\theta_{xx, \vec{r}} + \theta_{xx, \vec{r} + \hat{y} } +
\theta_{xx, \vec{r} + \hat{z}} + \theta_{xx, \vec{r} + \hat{y} +
\hat{z}} \cr\cr & - & \theta_{xy, \vec{r} + \frac{\hat{x}}{2} +
\frac{\hat{y}}{2} } - \theta_{xy, \vec{r} - \frac{\hat{x}}{2} +
\frac{\hat{y}}{2} } \cr\cr & - & \theta_{xy, \vec{r} + \hat{z} +
\frac{\hat{x}}{2} + \frac{\hat{y}}{2} } - \theta_{xy, \vec{r} +
\hat{z} - \frac{\hat{x}}{2} + \frac{\hat{y}}{2} } \cr\cr & - &
\theta_{xz, \vec{r} + \frac{\hat{x}}{2} + \frac{\hat{z}}{2} } -
\theta_{xz, \vec{r} - \frac{\hat{x}}{2} + \frac{\hat{z}}{2} }
\cr\cr & - & \theta_{xz, \vec{r} + \hat{y} + \frac{\hat{x}}{2} +
\frac{\hat{z}}{2} } - \theta_{xz, \vec{r} + \hat{y} -
\frac{\hat{x}}{2} + \frac{\hat{z}}{2} } ) = \mathcal{B}_{yz}.
\eeqn The tensor field defined this way satisfies $\sum_i \nabla_i
\mathcal{B}_{ij} = 0$. The relation between $\mathcal{E}_{ij}$ and
$h_{ij}$ is the same as that between $\mathcal{B}_{ij}$ and
$A_{ij}$.

The dual variable $\tilde{\pi}_{ii}$ is defined on the dual sites,
which are the centers of the cubes, and $\tilde{\pi}_{ij}$ with $i
\neq j$ is located on the links $\vec{r} + \frac{\hat{k}}{2}$,
with $k \neq i$, $k\neq j$. $\tilde{\pi}_{xx}$ on the dual site
$\vec{r} + \frac{\hat{x}}{2} + \frac{\hat{y}}{2} +
\frac{\hat{z}}{2} $ is defined as  \beqn 2\tilde{\pi}_{xx, \vec{r}
+ \frac{\hat{x}}{2} + \frac{\hat{y}}{2} + \frac{\hat{z}}{2}  } = -
\mathcal{B}_{xz, \vec{r} + \hat{y} + \frac{\hat{x}}{2} +
\frac{\hat{z}}{2}} + \mathcal{B}_{xz, \vec{r} + \frac{\hat{x}}{2}
+ \frac{\hat{z}}{2}} \cr\cr + \mathcal{B}_{xy, \vec{r} + \hat{z} +
\frac{\hat{x}}{2} + \frac{\hat{y}}{2}} - \mathcal{B}_{xy, \vec{r}
+ \frac{\hat{x}}{2} + \frac{\hat{y}}{2}} = \mathcal{C}_{xx}. \eeqn
$\tilde{\pi}_{xy}$ on the link $\vec{r} + \frac{\hat{z}}{2}$ is
defined as \beqn 2\tilde{\pi}_{xy, \vec{r} + \frac{\hat{z}}{2}}
&=& - 2\mathcal{B}_{yz, \vec{r} + \frac{\hat{y}}{2} +
\frac{\hat{z}}{2}} + 2\mathcal{B}_{yz, \vec{r} - \frac{\hat{y}}{2}
+ \frac{\hat{z}}{2}} \cr\cr &+& \mathcal{B}_{yy, \vec{r} +
\hat{z}} - \mathcal{B}_{yy, \vec{r}} - \mathcal{B}_{zz, \vec{r} +
\hat{z}} \cr\cr &+& \mathcal{B}_{zz, \vec{r}} - \mathcal{B}_{xx,
\vec{r} + \hat{z}} + \mathcal{B}_{xx, \vec{r}} =
2\mathcal{C}_{xy}.\eeqn $\mathcal{C}_{ij}$ defined this way is
symmetric, traceless, and covariantly constant. The relation
between $\mathcal{E}_{ij}$ and the dual variable $\tilde{h}_{ij}$ is
identical to that between $\tilde{\pi}_{ij}$ and $A_{ij}$, after
exchanging sites with cubes, and plaquettes with links. On the lattice,
the divergence of $\mathcal{C}_{ij}$ reads \beqn \sum_i\nabla_i
\mathcal{C}_{ix} &=& \mathcal{C}_{xx, \vec{r} + \frac{\hat{x}}{2}
+ \frac{\hat{y}}{2} + \frac{\hat{z}}{2}} - \mathcal{C}_{xx,
\vec{r} - \frac{\hat{x}}{2} + \frac{\hat{y}}{2} +
\frac{\hat{z}}{2}} \cr\cr &+& \mathcal{C}_{xy, \vec{r} + \hat{y} +
\frac{\hat{z}}{2}} - \mathcal{C}_{xy, \vec{r} + \frac{\hat{z}}{2}}
\cr\cr &+& \mathcal{C}_{xz, \vec{r} + \hat{z} + \frac{\hat{y}}{2}}
- \mathcal{C}_{xz, \vec{r} + \frac{\hat{y}}{2}}. \eeqn

\bibliography{emeb}

\end{document}